\documentclass[11pt]{article}
\usepackage{graphicx}
\usepackage{graphics}
\usepackage{amsfonts}
\newcommand{\R}{\mathbb{R}}
\newcommand{\Z}{\mathbb{Z}}
\newcommand{\Q}{\mathbb{Q}}

\begin{document}
\title{The Haar Wavelet Transform of a Dendrogram}
\author{Fionn Murtagh\thanks{Department of Computer Science, 
Royal Holloway, University of London, Egham TW20 0EX, England.  
{\tt fmurtagh@acm.org}}
}
\markboth{Haar Wavelet Transform of a Dendrogram}{Haar Wavelet 
Transform of a Dendrogram}
\maketitle

\begin{abstract}
We describe a new wavelet transform, for use on hierarchies or binary 
rooted trees.  The theoretical framework of this approach to data 
analysis is described.  Case studies are used to further exemplify this
approach.  A first set of application studies  deals with data array 
smoothing, or filtering.  A second set of application studies relates to 
hierarchical tree condensation.  Finally, a third study explores the 
wavelet decomposition, and the  reproducibility of data sets such as 
text, including a new perspective on the generation or computability of
such data objects.  
\end{abstract}


\section{Introduction}

In this paper, the new data analysis approach to be described can be 
understood as 
a transform which maps a hierarchical clustering into a transformed set of 
data; and this transform is invertible, meaning that the original data can 
be exactly reconstructed.  Such transforms are very often used in data 
analysis and signal processing because processing of the data may be 
facilitated by carrying out such processing in transform space, followed
by reconstruction of the data in some ``good approximation'' sense.  

Consider data smoothing as a case in point of such processing.  
Smoothing of data is important for exploratory visualization, 
for data understanding and interpretation, and as an aid 
in  model fitting (e.g., in time series analysis or more generally in 
regression modeling).  The wavelet transform is often used for signal (and
image) smoothing in view of its ``energy compaction'' properties, i.e., 
large values tend to become larger, and small values smaller, when the 
wavelet transform is applied.  Thus a very effective approach to signal 
smoothing is to selectively modify wavelet coefficients (for example, put 
small wavelet coefficients to zero) before reconstructing an approximate 
version of the data.  See H\"ardle (2000), Starck and Murtagh (2006).

The wavelet transform, developed for signal and image processing, has been 
extended for use on relational data tables and multidimensional data
sets (Vitter and Wang, 1999; Joe, Whang and Kim, 2001) 
for data summarization (micro-aggregation) 
with the goal of anonymization (or statistical disclosure limitation) 
 and macrodata generation; and data 
summarization with the goal of computational efficiency, especially in query 
optimization.  A survey of data mining applications (including 
applications to image and signal content-based information retrieval) 
can be found in Tao Li, Qi Li, Shenghuo Zhu and Ogihara (2002). 

A hierarchical representation is used by us, as a first phase of the 
processing, (i) in order to cater for the lack of any inherent row/column 
order in the given data table and to get around this obstacle to freely 
using a wavelet transform; and (ii) to take into account structure and
interrelationships in the data.  For the latter, a hierarchical clustering
furnishes an embedded set of clusters, and obviates any need for a priori 
fixing of number of clusters.  
Once this is done, the hierarchy is wavelet transformed.  The approach 
is a natural and integral one.

Our
innovation is to apply the Haar wavelet transform to a binary rooted
tree (viz., the clustering hierarchy) in terms of the following algorithm:
recursively carry out pairwise averaging and differencing at the sequence
of levels in the tree.  

A hierarchy may be constructed 
through use of any constructive, hierarchical clustering algorithm
(Benz\'ecri, 1979; Johnson, 1967; Murtagh, 1985). 
In this work we will assume that some
agglomerative criterion is satisfactory from the perspective of the type of 
data, and the nature of the data analysis or processing.  In a wide range of
practical scenarios, the minimum variance (or Ward) agglomerative 
criterion can be strongly recommended due to its data summarizing properties 
(Murtagh, 1985).

The remainder of this article is organized as follows.  
Sections \ref{sect3} and \ref{sect4} present important background 
context.  
Section \ref{sect5} presents our new wavelet transform.
In section \ref{sect6}, illustrative case studies are used to 
further discuss the new approach. 
Section \ref{sect7} deals with the application to data array smoothing,
or filtering.
Section \ref{sectcollapse} deals with the application to hierarchical 
tree condensation.
Section \ref{sect9} explores the wavelet decomposition, and linkages with 
reproducibility or recreation of data sets such as text.  

\section{Wavelets on Local Fields}
\label{sect3}

Wavelet transform analysis is the determining of a ``useful''
basis for $L^2(\R^m)$ which is induced from a discrete subgroup of $\R^m$,
and uses translations on this subgroup, and dilations of the basis functions.

Classically (Frazier, 1999; Debnath and Mikusi\'nski, 1999;
Strang and Nguyen, 1996)
the wavelet transform avails of a wavelet function $\psi(x) \in
L^2(\R)$, where the latter is the space of all square integrable functions
on the reals.
Wavelet transforms are bases on $L^2(\R^m)$, and the discrete lattice
subgroup $\Z^m$ ($m$-dimensional integers)
is used to allow discrete groups of dilated translation operators
to be induced on $\R^m$.  Discrete lattice subgroups are typical of
2D images (where the lattice is a pixelated grid) or 3D images 
(where the lattice is
a voxelated grid) or spectra or time series (the lattice is the set of
time steps, or wavelength steps).

Sometimes it is appropriate to consider the construction of wavelet
bases on $L^2(G)$ where $G$ is some group other than $\R$.  In
Foote, Mirchandani, Rockmore, Healy and Olson (2000a, 2000b; see also
Foote, 2005) this is done
for the group defined by a  quadtree, in turn derived from a 2D image.
To consider the wavelet transform approach not in a Hilbert space
but rather in locally-defined and discrete spaces we have to change the
specification of a wavelet function in $L^2(\R)$ and instead use
$L^2(G)$.

Benedetto (2004) and Benedetto and Benedetto (2004) considered in detail
the group $G$ as a locally compact abelian group.
Analogous to the integer grid, $\Z^m$, a compact subgroup is used to allow a
discrete group of operators to be defined on $L^2(G)$.
The property of locally compact (essentially: finite and free of edges)
abelian (viz., commutative) groups that is most important is the
existence of the Haar measure.  The Haar measure allows
integration, and definition of a topology on the algebraic structure of
the group.

Among the cases of wavelet bases constructed via a sub-structure 
are the following (Benedetto, 2004).

\begin{itemize}
\item Wavelet basis on $L^2(\R^m)$ using translation operators defined
on the discrete lattice, $\Z^m$.  This is the situation that 
holds for image processing, signal processing, most time series
analysis (i.e., with equal length time steps), spectral signal processing,
and so on.  As pointed out by Foote (2005), this framework allows the
multiresolution analysis in $L^2(\R^m)$ to be generalized to $L^p(\R^m)$
for Minkowski metric $L^p$ other than Euclidean $L^2$. % p. 1 of Foote '05

\item Wavelet basis on $L^2(\Q_p)$, where $\Q_p$ is the p-adic field,
using a discrete set of translation operators.  This case has been
studied by Kozyrev, 2002, 2004; Altaisky, 2004, 2005.  See also the 
interesting overview of Khrennikov and Kozyrev (2006).

\item Finally the central theme of Benedetto (2004) is a wavelet basis
on $L^2(G)$ where $G$ is a locally compact abelian group, using translation
operators defined on a compact open subgroup (or operators that can be used
as such on a compact open subgroup); and with definition of an
expansive automorphism replacing the traditional use of dilation.
\end{itemize}

In Murtagh (2006) the latter theme is explored: a p-adic representation of 
a dendrogram is used, and an expansive operator is defined, which, when
applied to a level of a dendrogram  enables movement up a level.  

In our case we are looking for a new basis for $L^2(G)$ where $G$ is the
set of all equivalent representations of a hierarchy, $H$, on $n$ terminals.
Denoting the level index of $H$ as $\nu$ (so $\nu : H \longrightarrow \R^+$,
where $\R^+$ are the positive reals), and $\nu = 0$ is the level index
corresponding to the fine partition of singletons, then this hierarchy will
also be denoted as $H_{\nu = 0}$.  Let $I$ be the set of observations.
Let the succession of clusters associated with nodes in $H$ be denoted
$Q = \{ q_1, q_2, \dots , q_{n-1} \}$.
We have $n-1$ non-singleton nodes in $H$, associated with the clusters, $q$.
At each node we can interchange left and right subnodes.  Hence we have
$2^{n-1}$ equivalent representations of $H$, or, again, members in the
group, $G$, that  we are considering.

So we have the group of equivalent dendrogram representations on
$H_{\nu = 0}$.  We have a series of subgroups, $H_{{\nu}_k} \supset
H_{{\nu}_{(k+1)}}$, for $0 \leq k <  n-1$.
Symmetries (in the group sense)
are given by permutations at each level, $\nu$, of hierarchy $H$.
Collecting these furnishes a group of symmetries on the terminal set of any
given (non-terminal) node in $H$.  

We want to process dendrograms, and we want our processing to be
invariant relative to any equivalent representation of a given dendrogram.

 Denote the permutation at level
$\nu$ by $P_\nu$.  Then the automorphism group is given by:
$$G = P_{n-1} \ \mathrm{wr} \ P_{n-2} \ \mathrm{wr} \ \dots \ \mathrm{wr} 
\ P_2 \  \mathrm{wr} \  P_1$$
where wr denotes the wreath product.  

Foote et al.\ (2000a, 200b) and Foote (2005) consider the wreath product
group of a tree representation of data, including the quadtree which is a 
tree representation of an image.  Just as for us here, the offspring nodes
of any given node in such a tree can be ``rotated'' ad lib.  Group action 
amounts to
cyclic shifts or adjacency-preserving permutations of the offspring nodes.
The group in this case is referred to as the wreath product group.

We will introduce and study a wavelet transform on $L(G)$ where $G$ is
the wreath product group based on the hierarchy or rooted binary tree, $H$.  

\section{Hierarchy, Binary Tree, and Ultrametric Topology}
\label{sect4}

A short set of definitions follow, showing how a hierarchy is taken 
in the form of a binary tree, and the particular form of binary tree used
here is often termed a dendrogram.  By small abuse of terminology, 
we will use $H$ to denote this hierarchy, and the ultrametric topology that
it represents.  

A hierarchy, $H$,
is defined as a binary, rooted, unlabeled, node-ranked tree, also
termed a dendrogram (Benz\'ecri, 1979; Johnson, 1967; Lerman, 1981; Murtagh,
1985).
A hierarchy defines a set of embedded subsets of a given set, $I$.  However
these subsets are totally ordered by an index function $\nu$, which is a
stronger condition than the partial order required by the subset relation.
A bijection exists between a hierarchy and an ultrametric space.

Let us show these equivalences between embedded subsets, hierarchy, and
binary tree, through the constructive approach of inducing $H$ on a set
$I$.

Hierarchical agglomeration on $n$ observation vectors, $i \in I$, involves
a series of $1, 2, \dots , n-1$ pairwise agglomerations of
observations or clusters, with the following properties.  A hierarchy
$H = \{ q | q \in 2^I \} $ such that (i) $I \in H$, (ii) $i \in H \ \forall
i$, and (iii) for each $q \in H, q^\prime \in H: q \cap q^\prime \neq
\emptyset \Longrightarrow q \subset  q^\prime \mbox{ or }  q^\prime
 \subset q$.  Here we have denoted the power set of set $I$ by $2^I$.
An indexed hierarchy is the pair $(H, \nu)$ where the positive
function defined on $H$, i.e., $\nu : H \rightarrow \R^+$, satisfies:
$\nu(i) = 0$ if $i \in H$ is a singleton; and (ii)  $q \subset  q^\prime
\Longrightarrow \nu(q) < \nu(q^\prime)$.  Here we have denoted the
positive reals, including 0, by $\R^+$.
Function $\nu$ is the agglomeration
level.  Take  $q \subset  q^\prime$, let $q \subset q''$
 and $q^\prime \subset q''$, and let $q''$ be the lowest level cluster for
which this is true. Then if we define $D(q, q^\prime) = \nu(q'')$, $D$ is
an ultrametric.  In practice, we start with a Euclidean or other
dissimilarity, use some criterion such as minimizing the change in variance
resulting from the agglomerations, and then define $\nu(q)$ as the
dissimilarity associated with the  agglomeration carried out.

\section{The Hierarchic Haar Wavelet Transform Algorithm: Description}
\label{sect5}

Linkages between the classical wavelet transform, as used in signal 
processing, and multivariate data analysis, were investigated in 
Murtagh (1998).  The wavelet transform to be described now is 
fundamentally different, and works on a hierarchy. 

The traditional 
Haar wavelet transform can be simply described in terms of the 
following algorithm: recursively carry out averaging and differencing 
of adjacent pairs of data values (pixels, voxels, time steps, etc.) at a 
sequence of geometrically (factor 2) increasing resolution levels.  
As mentioned in the Introduction, our 
innovation is to apply the Haar wavelet transform to 
the clustering hierarchy, and this algorithm is the 
recursive carrying  out of pairwise averaging and differencing at the sequence
of levels in the hierarchical tree.  

\subsection{Short Description of the Algorithm}

A dendrogram on $n$ terminal nodes, associated with observation vectors,
has $n-1$ non-terminal nodes.  We proceed through each of these non-terminal
nodes in turn, starting at the node corresponding to the sequentially 
first agglomeration, continuing to the node corresponding to the 
sequentially second agglomeration, and so on, until we finally reach the 
root node.  At each node, we define a vector as the (unweighted) 
average of the vectors of its two child nodes.  So the vector associated
with the very first non-terminal node will be the (unweighted) average
of the vectors comprising this node's two (terminal) child nodes.  

For subsequent non-terminal nodes, their child nodes may be terminal
or non-terminal.
In all cases, this procedure is well-defined.  We continue the 
procedure until we have processed all $n-1$ non-terminal nodes.  

We now have an increasingly smooth vector corresponding to each node in 
the dendrogram, or hierarchy, $H$.  We term this vector at each node the 
{\em smooth signal} or just {\em smooth} at each node.  

The {\em detail signal}, or {\em detail}, at each node is 
defined as the {\em vector difference} 
between the vector at a (non-terminal) node, and the vector at its 
(terminal or non-terminal) child node.  By consistently labeling 
left and right child subnodes, by construction the left child subnode 
will have a detail vector which is just the negative of the detail vector of 
the right subnode.  Hence, with consistency of left and right labeling,
we just need to store one of these detail vectors.

Because of the way that the detail signal has been defined, and given the 
smooth signal associated with the root node (the node with sequence number 
$n-1$), we can easily see the following: to reconstruct the original data,
used to set this algorithm underway, we need just the set of all detail 
signals, and the final, or root node, smooth signal.  

\subsection{Definition of Smooth Signals and Detail Signals}

Consider any hierarchical clustering, $H$, represented as a binary rooted
tree. For each cluster associated with a non-terminal node, 
$q''$, with offspring (terminal or non-terminal) 
nodes $q$ and $q'$, we define $s(q'')$ through 
application of the low-pass filter 
$\left(
\begin{array}{r}
 \frac{1}{2} \\  
\frac{1}{2} 
\end{array}
\right)
$ which can be implemented as a scalar product:

\begin{equation}
s(q'') = \frac{1}{2} \left( s(q) + s(q') \right) = 
\left(
\begin{array}{r}
0.5 \\
0.5
\end{array}
\right)^t
\left(
\begin{array}{c}
s(q) \\
s(q')        
\end{array}
\right)
\label{eqn1}
\end{equation}

The application of the low-pass filter is carried out in order of 
increasing node number (i.e., from the smallest non-terminal node, 
through to the root node).  For a terminal node, $i$, allowing us to 
notationally say that $q = i$ or that $q' = i$, the signal smooth 
$s(i)$ is just the 
given vector, and this aspect is addressed further below, in subsection
\ref{twocases0}.

Next for each cluster
$q''$ with offspring nodes $q$ and $q'$, we define detail coefficients 
$d(q'')$ through 
application of the band-pass filter $\left( 
\begin{array}{r}
\frac{1}{2} \\ 
-\frac{1}{2} 
\end{array}
\right)$: 

\begin{equation}
d(q'') = \frac{1}{2} ( s(q) - s(q') ) = 
\left(
\begin{array}{r}
0.5 \\
-0.5
\end{array}
\right)^t
\left(
\begin{array}{c}
s(q) \\
s(q')
\end{array}
\right)
\label{eqn2}
\end{equation}

Again, increasing order of node number is used for application of this
filter.

\begin{figure*}
\begin{center}
\includegraphics[width=14cm,angle=270]{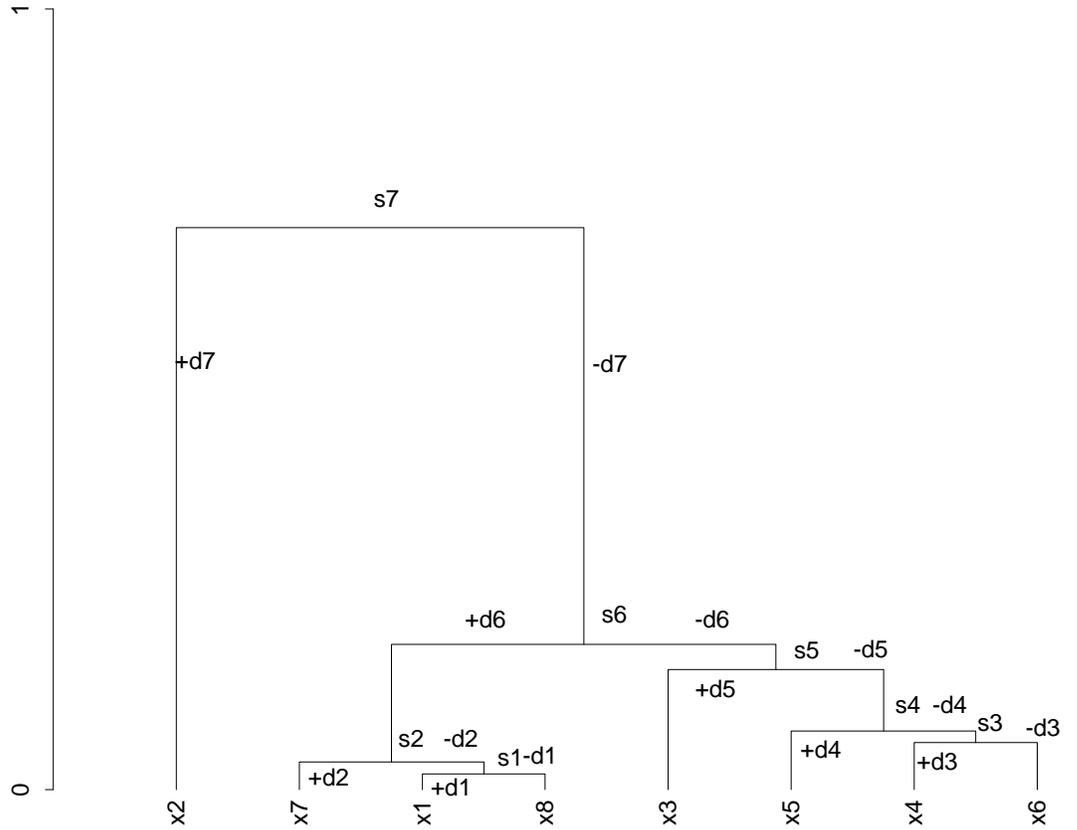}
\end{center}
\caption{Dendrogram on 8 terminal nodes constructed from first 8 values
of Fisher iris data.  (Median agglomerative method used in this case.)
Detail or wavelet coefficients are denoted by $d$, and data smooths are 
denoted by $s$.  The observation vectors are denoted by $x$ and are
associated with the terminal nodes.  
Each {\em signal smooth}, $s$, is a vector.  The (positive
or negative) {\em detail signals}, $d$, are also vectors.  All these 
vectors are of the same dimensionality.}
\label{fig5}
\end{figure*}

The scheme followed is illustrated in Figure \ref{fig5}, which shows the
hierarchy (constructed by the median agglomerative method, although this 
plays no role here), using for display convenience just the first 
8 observation vectors in Fisher's iris data (Fisher, 1936).

We call our algorithm a Haar wavelet transform because, traditionally,
this wavelet transform is defined by a similar set of averages and 
differences.   The former, low-pass filter, is used to set the center 
of the two clusters being agglomerated; and the latter, band-pass filter,
is used to set the deviation or discrepancy of these two clusters from the 
center.  

\subsection{The Input Data}
\label{twocases0}

We now return to the issue of how we start this scheme, i.e.\ how we 
define $s(i)$, or the ``smooth'' of a terminal node, representing a 
singleton cluster.  

Let us consider two cases: 

\begin{enumerate}
\item $s(i) $ is a vector in $\R^m$, and the $i$th row of a data table.
\item $s(i) $ is an $n$-dimensional indicator vector.  So the third,
in sequence, out of a population of $n = 8$ observations has indicator 
vector $\{00100000\}$.  We can of course take a data table 
of all indicator vectors: it is clear that the data table is symmetric, and
is none other than the identity matrix.  
\end{enumerate}

Our hierarchical Haar wavelet transform can easily handle either case, 
depending on the input data table used.  

While we have considered two cases of input data, we may note the following.
Having the clustering hierarchy built on the same input data 
as used for the hierarchical Haar transform is reasonable when compression of
the input data is our target.  However, if the hierarchical Haar transform 
is used for data approximation, cf.\ section \ref{sect9} below, then we are
at liberty to use different data for the hierarchical clustering and for 
the Haar transform.  The hierarchy is built from a set of observation 
vectors in $\R^m$.  Then it is used in the Haar wavelet transform as a 
structure on another set of vectors in $\R^{m'}$ (with $m'$ not 
necessarily equal to $m$).  
We will not pursue this line of investigation further 
here.  

\subsection{The Inverse Transform}

Constructing the hierarchical Haar wavelet transformed data is referred to 
as the forward transform.  Reconstructing the input data is 
referred to as the inverse transform.  

The inverse transform allows exact reconstruction of the input data.
We begin with $s_{n-1}$.  If this root node has subnodes $q$ and $q'$, we
use $d(q)$ and $d(q')$ to form $s(q)$ and $s(q')$.

We continue, step by step, until we have reconstructed all vectors
associated with terminal nodes.

\subsection{Matrix Representation}
\label{sect45}

Let our input data be a set of $n$ points in $\R^m$ given in the form of
matrix $X$.  We have:

\begin{equation}
X = C D + S_{n-1}
\end{equation}

\noindent
where $D$ is the matrix collecting all
 wavelet projections or detail coefficients, $d$.
The dimensions of $D$ are $(n-1) \times m$.
The dimensions of $C$ are: $n \times (n-1)$.  $C$ is a characteristic 
matrix representing the dendrogram.  Murtagh (2006) provides an 
introduction to $C$, which will be summarized here.  In Figure \ref{fig2},
the 0 or 1 coding works well when we also take account of the existence of 
the node at that particular level.  Other forms of coding can be used, and 
Murtagh (2006) uses a ternary code, viz., $+1$ and $-1$ for left and
right branches (replacing 0 and 1 in Figure \ref{fig2}, respectively), and 
0 to indicate non-existence of a node at that particular level.  

Matrix $C$, describing the branching codes $+1$ and $-1$,
and an absent or non-existent branching given by $0$,
uses a
set of values $c_{ij}$ where $i \in I$, the indices of the object set; and
$j \in \{ 1, 2, \dots , n-1 \}$, the indices of the dendrogram levels
or nodes ordered
increasingly.  For Figure \ref{fig2} we  have:

\begin{equation}
C = \{ c_{ij} \} =
\left(
\begin{array}{rrrrrrr}
1  & 1 & 0 &  0 & 1 &  0 & 1 \\
-1 & 1 & 0 &  0 & 1 & 0 & 1 \\
0  & -1 & 0 &  0 & 1 & 0 & 1 \\
0  &  0 & 1 &  1 & -1 & 0 & 1 \\
0  & 0  & -1 & 1 & -1 & 0 & 1 \\
0  & 0 & 0  & -1 & -1 & 0 & 1 \\
0  & 0 & 0  & 0  & 0 & 1 & -1 \\
0  & 0 & 0  & 0  & 0 & -1 & -1
\end{array}
\right)
\label{eqn222}
\end{equation}

For given level $j$, $\forall i$, the absolute values $| c_{ij} |$
give the membership function either by node, $j$, which is therefore
read off
columnwise; or by object index, $i$ which is therefore read off rowwise.

If $s_{n-1}$ is the final data smooth, in the limit for very large $n$
a constant-valued $m$-component vector, then let $S_{n-1}$ be
the $n \times m$
matrix with $s_{n-1}$ repeated on each of the $n$ rows.

Consider the $j$th coordinate of the $m$-dimensional observation
vector corresponding to $i$.
For any $d(q_j)$ we have: $ \sum_k d(q_j)_k = 0 $, i.e.\ the detail
coefficient vectors are each of zero mean.

In the case where out input data consists of $n$-dimensional 
indicator vectors (i.e., the $i$th vector contains 0-values except for
location $i$ which has a 1-value), then our  
initial data matrix $X$ is none other than the $n \times n$
dimensional identity matrix.  We will write $X_{\mbox{ind}}$ for this
identity matrix.

The wavelet transform in this case is: $X_{\mbox{ind}} = C D + S_{n-1}$.

$X_{\mbox{ind}}$ is of dimensions $ n \times n$.

$C$, exactly as before, is a characteristic matrix representing the 
dendrogram used, and is of dimensions $ n \times (n-1)$.

$D$, of necessity different in values from case 1, is of dimensions
$ (n-1) \times n$.

$S_{n-1}$, of necessity different in values from case 1, is of dimensions
$ n \times n$.

\subsection{Computational Complexity Properties}

The computational complexity of our algorithms are as follows.  The 
hierarchical clustering is $O(n^2)$.
The forward hierarchical
Haar wavelet transform is $O(n)$.  Finally, 
the inverse wavelet transform is $O(n^2)$.  
On Macintosh G4 or G5 machines, all phases of the 
processing took typically 4--5 minutes for an array of dimensions
 $12000 \times 400$.  

An exemplary pipeline of C and R code used in this work 
is available at the following address: 
http://astro.u-strasbg.fr/$\sim$fmurtagh/mda-sw

\section{Hierarchical Haar Wavelet Transform: Case Studies}
\label{sect6}

In a practical way, using small data sets, 
we will describe our new hierarchical Haar wavelet 
transform in this section.  

\begin{figure*}
\begin{center}
\includegraphics[width=14cm,angle=270]{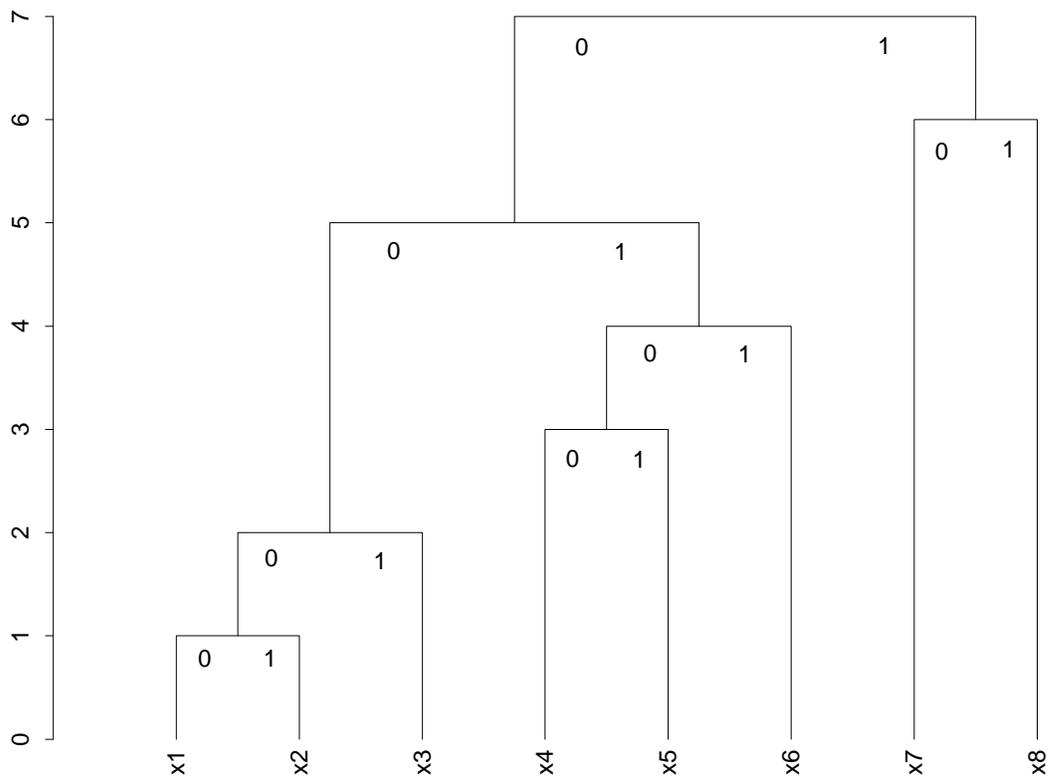}
\end{center}
\caption{Labeled, ranked dendrogram on 8 terminal nodes.  Branches labeled 
0 and 1.}
\label{fig2}
\end{figure*}

Let us begin with the indicator vectors case.  Thus in Figure 
\ref{fig2}, $x_1 = \{ 100000000 \}$, and $q_1 = 
\{ 11000000 \}$.  Note that this is our definition of $x_1$ etc. (and 
is not read off the tree), and from the definition of $x_1$ and $x_2$ we
have defined $q_1$.   This form of coding was used by Nabben and Varga (1994).

Now we use equations \ref{eqn1} and \ref{eqn2}.

\subsection{Case Study 1}

In Figure \ref{fig2}, we have:

\medskip

$ s(q_1) = \frac{1}{2} ( x_1 + x_2 ) = 
( \frac{1}{2} \ \frac{1}{2} \ 0 \ 0 \ 0 \ 0 \ 0 \ 0 ) $

Also: $s(q_1) = \frac{1}{2} q_1 $

\medskip

$ s(q_2) = \frac{1}{2} ( s(q_1) + x_3 ) = 
(\frac{1}{4} \ \frac{1}{4} \ \frac{1}{2} \ 0 \ 0 \ 0 \ 0 \ 0 )$

Also: $s(q_2) = \frac{1}{2} ( \frac{1}{2} q_1 + x_3 )  = \frac{1}{4} q_1 + 
\frac{1}{2} x_3 $

\medskip

$ s(q_3) = \frac{1}{2} ( x_4 + x_5 ) = 
( 0 \ 0 \ 0 \ \frac{1}{2} \ \frac{1}{2} \ 0 \ 0 \ 0 ) $

Also: $s(q_3) = \frac{1}{2} q_3 $

\medskip

$s(q_4) = \frac{1}{2} ( s(q_3) + x_6 ) = 
(0 \ 0 \ 0 \ \frac{1}{4} \ \frac{1}{4} \ \frac{1}{2} \ 0 \ 0)$

\medskip

$s(q_5) = \frac{1}{2} ( s(q_2) + s(q_4) ) = 
(\frac{1}{8} \ \frac{1}{8} \ \frac{1}{4} \ \frac{1}{8} \ \frac{1}{8} 
\frac{1}{4} \ 0 \ 0 )$

\medskip

$s(q_6) = \frac{1}{2} ( x_7 + x_8 ) = ( 0 \ 0 \ 0 \ 0 \ 0 \ 0 \ \frac{1}{2}
\ \frac{1}{2} ) $

\medskip

$s(q_7) = \frac{1}{2} ( q_5 + q_6 ) = ( \frac{1}{16} \ \frac{1}{16} \ 
\frac{1}{8} \ \frac{1}{16} \ \frac{1}{16} \ \frac{1}{8} \ \frac{1}{4} \
\frac{1}{4} ) $

\medskip
Next we turn attention to the detail coefficients.
\medskip

$ d(q_1) = \frac{1}{2} ( x_1 - x_2 ) = 
( \frac{1}{2} \ -\frac{1}{2} \ 0 \ 0 \ 0 \ 0 \ 0 \ 0 ) $

Alternatively, for $q'' = q \cup q'$, the detail coefficients are defined as:
$d(q'') = s(q'') - s(q') = - (s(q'') - s(q)) $.

\medskip

Thus $d(q_1) = s(q_1) - x_2 = ( \frac{1}{2} \ \frac{1}{2} \ 0 \ 0 \ 0 \ 0 \ 0
\ 0 ) - ( 0 \ 1 \ 0 \ 0 \ 0 \ 0 \ 0 \ 0 ) =
( \frac{1}{2} \ -\frac{1}{2} \ 0 \ 0 \ 0 \ 0 \ 0 \ 0 ) $

\medskip

For any $d(q_j)$ we have: $ \sum_k d(q_j)_k = 0 $, i.e.\ the detail 
coefficient vectors are each of zero mean.

Let us redo in vector and matrix terms this description of the 
hierarchical Haar wavelet transform algorithm.  

We take our initial or input data as follows.

\begin{equation}
\left(
\begin{array}{c}
x_1 \\
x_2 \\
x_3 \\
x_4 \\
x_5 \\
x_6 \\
x_7 \\
x_8 
\end{array}
\right)
=
\left(
\begin{array}{cccccccc}
1  & 0 & 0 & 0 & 0 & 0 & 0 & 0 \\
0  & 1 & 0 & 0 & 0 & 0 & 0 & 0 \\
0  & 0 & 1 & 0 & 0 & 0 & 0 & 0 \\
0  & 0 & 0 & 1 & 0 & 0 & 0 & 0 \\
0  & 0 & 0 & 0 & 1 & 0 & 0 & 0 \\
0  & 0 & 0 & 0 & 0 & 1 & 0 & 0 \\
0  & 0 & 0 & 0 & 0 & 0 & 1 & 0 \\
0  & 0 & 0 & 0 & 0 & 0 & 0 & 1 
\end{array}
\right)
\label{eqn5555}
\end{equation}

The hierarchical Haar wavelet transform of this input data is then as follows.

\begin{displaymath}  
\left(
\begin{array}{c}
d(q_1) \\
d(q_2) \\
d(q_3) \\
d(q_4) \\
d(q_5) \\
d(q_6) \\
d(q_7) \\
s_7 
\end{array}
\right)       =                                                
\end{displaymath}

\begin{equation}
\left(
\begin{array}{rrrrrrrr}
\frac{1}{2}  & -\frac{1}{2} & 0 & 0 & 0 & 0 & 0 & 0 \\
\frac{1}{4}  & \frac{1}{4} & -\frac{1}{2} & 0 & 0 & 0 & 0 & 0 \\
0  & 0 & 0 & \frac{1}{2} & -\frac{1}{2} & 0 & 0 & 0 \\
0  & 0 & 0 & \frac{1}{4} & \frac{1}{4} & -\frac{1}{2} & 0 & 0 \\
\frac{1}{8}  & \frac{1}{8} & \frac{1}{4} & -\frac{1}{8} & -\frac{1}{8} & 
   -\frac{1}{4} & 0 & 0 \\
0 & 0 & 0 & 0 & 0 & 0 & \frac{1}{2} & -\frac{1}{2} \\
\frac{1}{16}  & \frac{1}{16} & \frac{1}{8} & \frac{1}{16} & \frac{1}{16} & 
   \frac{1}{8} & -\frac{1}{4} & -\frac{1}{4} \\
\frac{1}{16}  & \frac{1}{16} & \frac{1}{8} & \frac{1}{16} & \frac{1}{16} & 
   \frac{1}{8} & \frac{1}{4} & \frac{1}{4} 
\end{array}
\right)
\label{eqn6}
\end{equation}

As already noted in this subsection, the succession of $n - 1$ wavelet
coefficient vectors are of zero mean.  Therefore, due to the input data
used (relation (\ref{eqn5555})), each row of the 
right hand matrix in equation \ref{eqn6} is of zero mean.

Note that this transform is a function of the hierarchy, $H$.  Here we are
using the hierarchy of Figure \ref{fig2}.  $H$ is needed to define the 
structure of the right hand matrix in equation \ref{eqn6}.  (This is 
closely related to the discussion in subsection \ref{sect45}.  Further 
discussion can be found in Murtagh, 2006).  

\subsection{Case Study 2}

In Tables \ref{table5} and \ref{table6} we directly transform a small
data set consisting of the first 8 observations in Fisher's iris data.

Note that in Table \ref{table6} it is entirely appropriate that at more 
smooth levels (i.e., as we proceed through levels d1, d2, $\dots$, d6, d7)
the values become more ``fractionated'' (i.e., there are 
more values after the decimal point).  

The minimum variance agglomeration criterion, with Euclidean
distance, is used to induce the hierarchy on the given data.  Each detail   
signal is of dimension $m = 4$ where $m$ is the dimensionality of the given 
data.  The smooth signal is of dimensionality $m$ also.  The number of 
detail or wavelet signal levels is given by the number of levels in the
labeled, ranked hierarchy, i.e.\ $n-1$.  

\begin{table}
\begin{center}
\begin{tabular}{|rrrrr|} \hline
    &   Sepal.L   &  Sepal.W    &  Petal.L  &   Petal.W \\ \hline
1   &       5.1   &      3.5    &      1.4  &       0.2 \\
2   &       4.9   &      3.0    &      1.4  &       0.2 \\
3   &       4.7   &      3.2    &      1.3  &       0.2 \\
4   &       4.6   &      3.1    &      1.5  &       0.2 \\
5   &       5.0   &      3.6    &      1.4  &       0.2 \\
6   &       5.4   &      3.9    &      1.7  &       0.4 \\
7   &       4.6   &      3.4    &      1.4  &       0.3 \\
8   &       5.0   &      3.4    &      1.5  &       0.2 \\ \hline
\end{tabular}
\end{center}
\caption{First 8 observations of Fisher's iris data.  L and W
refer to length and width.}
\label{table5}
\end{table}

\begin{table*}
\begin{center}
\begin{tabular}{|rrrrrrrrr|} \hline
        &     s7  &     d7    &    d6  &   d5  &   d4   &  d3  &  d2 &    d1 \\ 
\hline
Sepal.L & 5.146875 & 0.253125 & 0.13125 & 0.1375 &$-0.025$ & 0.05 & $-0.025$ &  
0.05 \\
Sepal.W & 3.603125 & 0.296875 & 0.16875 & $-0.1375$ & 0.125 & 0.05 & $-0.075$ & 
$-0.05$ \\
Petal.L & 1.562500 & 0.137500 & 0.02500    & 0.0000   & 0.000  & $-0.10$ & 0.050
 & 0.00 \\
Petal.W & 0.306250 & 0.093750 & $-0.01250$ & $-0.0250$ & 0.050 & 0.00    & 0.000
 & 0.00 \\ \hline
\end{tabular}
\end{center}
\caption{The hierarchical Haar wavelet transform resulting from use of the 
first 8 observations of Fisher's iris data shown in Table \ref{table5}.  
Wavelet coefficient levels are denoted 
d1 through d7, and the continuum or smooth component is denoted s7.}
\label{table6}
\end{table*}

\subsection{Traditional versus Hierarchical Haar Wavelet Transforms}

Consider a set of 8 input data objects, each of which is scalar: \\
$(64, 48, 16, 32, 56, 56, 48, 24)$.  A traditional Haar wavelet transform 
of this data can be quickly done, and gives: 
$(43,     -3,    16,     10,     8,    -8,     0,     12)$.  Here, the first
value is the final smooth, and the remaining values are the wavelet 
coefficients read off by a traversal from final smooth towards the 
input data values.  Showing the output in the same way, the hierarchical
Haar wavelet transform of the same data gives: 
$(40,     14,     6,     -6,    -4,     4,     0,      0)$.  

A little
reflection shows that the greater number of zeros in the hierarchical 
Haar wavelet transform is no accident.  In fact, with the following 
conditions:  10 different digits in the input data;
processing of an $n$-length string of digits; equal frequencies of digits
(if necessary, supported by a supposition of large $n$); 
and use of an unweighted average agglomerative criterion; we have the 
following. 

The hierarchy begins with nodes that agglomerate identical values,
$n/2$ of them; followed by agglomeration of the next round of $n/4$ 
values; followed by $n/8$ agglomerations of identical values; etc.  
So we have, in all, $n/2 + n/4 + n/8 + \dots + n/2^{n-1}$ identical value
agglomerations.  All of these will give rise to 0-valued {\em detail}
coefficients.  For suitable $n$, all save the very last round of 10 
agglomerations will give rise to non-0 valued {\em detail} coefficients.

This remarkable result points to the powerful data compression potential 
of the hierarchical Haar wavelet transform.  We must note though that this
rests on the dendrogram, and the computational requirements of the latter 
are not in any way bypassed.

\section{Hierarchical Wavelet Smoothing and Filtering}
\label{sect7}

Previous work on wavelet transforms of data tables includes 
Chakrabarti, Garofalakis, Rastogi and Shim (2001) and Vitter and Wang (1999). 
There are problems, however, in directly applying a wavelet transform
to a data table.  Essentially, a relational table (to use database
terminology; or matrix)
is treated in the same way as a 2-dimensional pixelated 
image, although
the former case is invariant under row and column permutation, whereas the 
latter case is not (Murtagh, Starck and Berry, 2000).
Therefore there are immediate problems related to 
non-uniqueness, and data order dependence.   
What if, however, one organizes the data such that adjacency has 
a meaning?  This implies that similarly-valued objects, and/or 
similarly-valued features, are close together.  This is what we do, 
using any hierarchical clustering algorithm.  

From a given input data array, the 
hierarchical wavalet transform creates an output data array with the property
of forcing similar values to be close together.  This transform is 
fully reversible.  The proximity of similar values, however, is with respect
to the hierarchical tree which was used in the forward transform, and of 
course comes into play also in the inverse transform.  Because similar 
values are close together, the compressibility of the transformed data 
is enhanced.  Separately, setting small values in the transformed data to 
zero allows for data smoothing or data ``filtering''.  

\subsection{The Smoothing Algorithm}

We use the following generic data analysis processing path, which is 
applicable to any input tabular data array of numerical values.  We assume
only that there are no missing values in the data array. 

\begin{enumerate}
\item Given a dissimilarity, induce a hierarchy 
on the set of observations.  (We generally use the Euclidean distance, and the 
minimum variance agglomerative hierarchical clustering criterion, in view
of the synoptic properties, Murtagh, 1985.  Additionally
the vectors used in the clustering can be weighted: we use identical weights
in this work.) 
\item Carry out a Haar wavelet transform on this hierarchy. 
This gives a tree-based compaction of energy (Starck and Murtagh, 
2006: large values
tend to become larger, and small values tend to become smaller)
in our data.  Filter the wavelet coefficients (i.e., carry out wavelet 
regression or smoothing, here using hard thresholding (see e.g.\ Starck 
and Murtagh, 2006) by setting small
wavelet coefficients to zero).  
\item Determine the inverse of the wavelet transform, in order to 
reconstruct an approximation to the original multidimensional data values.
\end{enumerate}

\subsection{Fisher Iris Data and Uniformly Distributed Values of
Same Array Dimensions}

In this first filtering 
study we use Fisher's iris data (Fisher, 1936), an array of dimensions
$150 \times 4$, in view of its well known characteristics.  
If $x_{ij}$ is a typical data value, then the energy of 
this data is $1/(nm) \sum_{ij} x_{ij}^2 = 15.8988$.  If we set wavelet 
coefficients to zero based on a hard threshold, then a very large number 
of coefficients may be set to zero with minor implications for approximation
of the input data by the filtered output.  (A hard threshold uses a 
step function, and can be 
counterposed to a soft threshold, using some other, monotonical increasing,
function.  A hard threshold, used in Table \ref{table1}, 
is straightforward and, in the absence of 
any further a priori information, the most reasonable choice.)  
The minimum variance hierarchical clustering method was used as the first
phase of the processing, followed by the second, wavelet transform, phase.  
Then followed wavelet coefficient truncation, and reconstruction or the
inverse transform.  We see that
a mean square error between input and output of value 0.1040 is the 
global  approximation quality, when nearly 98\% of 
wavelet coefficients are zero-valued.  

To further illustrate what is happening in this approximation by the 
wavelet filtered data, Table \ref{table7} shows the last 10 iris observations,
as given for input, and as filtered.  The numerical precisions shown are
as generated in the reconstruction, which explains why we show some values 
to 4 decimal places, and some to 6.

\begin{table}
\begin{center}
\begin{tabular}{|rrr|}\hline
Filt.\ threshold & \% coeffs.\ set to zero & mean square error \\ 
\hline
0  & 16.95  &  0  \\
0.1 &  70.13  & 0.0098 \\
0.2 & 91.95 & 0.0487 \\
0.3 & 97.15 & 0.0837 \\
0.4 & 97.82 & 0.1040 \\ \hline
\end{tabular}
\end{center}
\caption{Hierarchical Haar smoothing results for Fisher's 
$150 \times 4$ iris data.}
\label{table1}
\end{table}

\begin{table}
\begin{center}
\begin{footnotesize}
\begin{tabular}{|rrrrr|} \hline
      &   Sepal.L   &  Sepal.W    &  Petal.L  &   Petal.W \\ \hline
140   &       6.9   &      3.1     &     5.4  &       2.1 \\
141   &       6.7   &      3.1    &      5.6  &       2.4 \\
142   &       6.9   &      3.1    &      5.1  &       2.3 \\
143   &       5.8   &      2.7    &      5.1  &       1.9 \\
144   &       6.8   &      3.2    &      5.9  &       2.3 \\
145   &       6.7   &      3.3    &      5.7  &       2.5 \\
146   &       6.7   &      3.0    &      5.2  &       2.3 \\
147   &       6.3   &      2.5    &      5.0  &       1.9 \\
148   &       6.5   &      3.0    &      5.2  &       2.0 \\
149   &       6.2   &      3.4    &      5.4  &       2.3 \\
150   &       5.9   &      3.0    &      5.1  &       1.8 \\ \hline
      &             &             &           &           \\
140   &  6.739063  &  3.119824   &    5.4125  &  2.239258 \\
141   &  6.782813  &  3.307324   &    5.7250  &  2.564258 \\
142   &  6.839063  &  3.119824   &    5.1125  &  2.239258 \\
143   &  5.737500  &  2.808496   &    5.0000  &  2.039258 \\
144   &  6.782813  &  3.307324   &    5.8250  &  2.314258 \\
145   &  6.782813  &  3.307324   &    5.7250  &  2.564258 \\
146   &  6.639063  &  3.119824   &    5.1125  &  2.239258 \\
147   &  6.196875  &  2.480371   &    5.0000  &  1.964258 \\
148   &  6.364063  &  3.019824   &    5.2625  &  2.089258 \\
149   &  6.320313  &  3.307324   &    5.4750  &  2.439258 \\
150   &  5.937500  &  3.008496   &    5.1375  &  1.864258 \\ \hline
\end{tabular}
\end{footnotesize}
\end{center}
\caption{Last 10 values of input data, and of the approximation to these
based on the hierarchical Haar wavelet transform filtering with a 
hard threshold of 0.1, implying 70.13\% of the wavelet coefficients 
equaling 0.}
\label{table7}
\end{table}

To show that wavelet filtering is effective, we will next compare wavelet
filtering with direct filtering of the given data. 
By ``directly filtered'' we mean that we
processed the original data without recourse to a hierarchical clustering.
This is intended as a simple, default baseline with 
which we can compare our results.  

Taking the 
{\em original} Fisher data, we find the median value to be 3.2.  Putting
values less than this median value to 0, we find the MSE to be 2.154567,
i.e., implying a far less satisfactory fit to the data.  (Thresholding 
by using $<$ versus $\leq$ median had no effect.)  

\subsection{Uniform Realization of Same Dimensions as Fisher Data}

We next generated 
an array of dimensions $150 \times 4$ of uniformly distributed
random values on $[ 0, 7.9 ]$, where 7.9 was the maximum value in the
Fisher iris data.  The energy of this data set was 21.2097.  
 Results of filtering are shown in Table \ref{table2}.  The minimum 
variance hierarchical clustering
method was used.  Again good approximation properties are seen,
even if the compression is not as impressive as for the Fisher data.

\begin{table}
\begin{center}
\begin{tabular}{|rrr|}\hline
Filt.\ threshold & \% coeffs.\ set to zero & mean square error \\ 
\hline
0  & 0 &  0  \\
0.1 &  14.77  & 0.0022 \\
0.2 & 31.54 & 0.0249 \\
0.3 & 42.79 & 0.0622 \\
0.4 & 53.52 & 0.1261 \\ \hline
\end{tabular}
\end{center}
\caption{Hierarchical Haar filtering results for uniformly distributed
$150 \times 4$ data.}
\label{table2}
\end{table}

Uniformly distributed data coordinate values
are a taxing case, since such data are very unlike data with clear cluster
structures (as is the case for the Fisher iris data).   

In the next subsection we will further explore this aspect of internal 
structure in our data. 

\subsection{Inherent Clustering Structure in a Data Array: Implications
for Wavelet Filtering}
\label{lastsect}

First we show that the influence of numbers of rows or columns in our
data array is very minor in regard to the wavelet filtering. 

When a data set is inherently clustered (and possibly inherently 
hierarchically clustered) then the energy compaction properties of the 
wavelet transformed data ought to be correspondingly stronger.  We will 
show this through the processing of data sets containing cluster structure
relative to the processing of data sets containing uniformly distributed
values (and hence providing a baseline for no cluster structure).  

Firstly we verified that data set size is relatively unimportant in terms of 
wavelet-based smoothing.  We took artificially generated, uniformly 
distributed in [0, 1], random data matrices of dimensions: 
$500 \times 40$, $1000 \times 40$, $1500 \times 40$, and $2000 \times 40$.  
For each we applied a fixed threshold of 0.1 to the wavelet coefficients, 
setting values less than or equal to this threshold to 0, and retaining 
wavelet coefficient values above this threshold, 
before reconstructing the data.  
Then we checked mean square error between reconstructed data and the original
data.  For the four different data matrix dimensions, we found: 
0.463, 0.461, 0.465, 0.466.  (MSE used here was: $1/n \   
\sum_{i,j} (\hat{x}_{ij} - x_{ij})^2$, where $\hat{x}$ is the filtered
data array value, $n$ is the number of observations indexed by $i$; 
and $x$ is the input data value.)

From the clustering point of view, the foregoing data matrices are 
simply clouds of 500, 1000, 1500 and 2000 points in 40-dimensional real space,
or $\R^{40}$.  To check if space dimensionality could matter we checked 
the mean square error for a data matrix of uniformly distributed values with
dimensions $2000 \times 400$,  the mean 
square error was found to be 0.458.  (Compare this to the mean square error
of 0.466 for the $2000 \times 40$ data array, discussed in the previous
paragraph.  A constant 10 divisor was used, for comparability of results, 
for the $2000 \times 400$ data.)

We conclude that neither embedding spatial dimensionality, i.e., number of 
columns in the data matrix, nor also data set size as given by the number 
of rows, are inherent determinants  of the smoothing properties of our 
new method.

So what is important?  Clearly if the hierarchical clustering is pulling 
large clusters together, and facilitating the ``energy compaction'' 
properties of the wavelet transform, then what is important is 
clustering structure in our data.

\subsection{Compressibility of the Transformed Data}

\begin{figure*}
\begin{center}
\includegraphics[width=10cm]{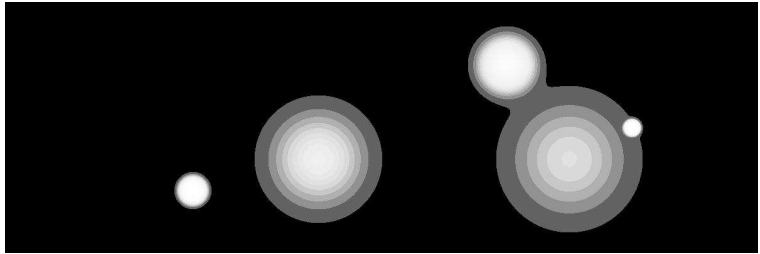}
\end{center}
\caption{Visualization of the artificial structure defined from 5 different 
Gaussian distributions, before noise was added.  
Data array dimensions: $1200 \times 400$ (portrayed transposed here).}
\label{figsim}
\end{figure*}

We generated structure by placing Gaussians centered at the following 
row, column locations in a $1200 \times 400$ data array:  300,100; 
800,300; 1000,200; 500,150; 900,150.  These bivariate Gaussians were of 
total 10 units in each case.  A full width at half maximum (equal 
to 2.35482 times the standard deviation of a Gaussian), 
was used in each case, respectively: 20, 50, 10, 
100, 125.  We will call this the data array containing structure. 
Figure \ref{figsim} shows a schematic view of it.  Next we added uniformly
distributed random noise, where the values in the 
former (Gaussians) data set scaled down by a constant factor of 10.  
Figure \ref{figsim2} shows the data set we will now work on.   
(We added noise in order to have a non-trivial data set for our 
compressibility experiments.)

\begin{figure*}
\begin{center}
LARGE FIGURE NOT DISPLAYED: SEE COPY OF PAPER AT 
http://www.cs.rhul.ac.uk/home/fionn/papers
\end{center}
\caption{Data array dimensions: $1200 \times 400$ (portrayed transposed).  
Structure is visible, 
and added uniform nose.}
\label{figsim2}
\end{figure*}

In Murtagh et al.\ (2000) it was noted how row and column permuting of a 
data table allows application of any wavelet transform, where we take the 
data table as a 2-dimensional image.  For a 3-way data array, a wavelet 
working on a 3-dimensional image volume is directly applicable.  The difficulty
in using an image processing technique on a data array is that (i) we must
optimally permute data table rows and columns, which is known to be an 
NP-complete problem, or (ii) we must accept that each alternative data table
row and column permutation will lead to a different result.  

To exemplify this situation, we take the data generated as described 
above, which is by construction optimally row/column permuted.  
Figure \ref{figsim2} shows the data used, generated as having structure 
(i.e., contiguous similar values, which can still be visually appreciated) 
but also 
subject to additional uniformly distributed noise.  The noise notwithstanding,
there is still structure present in this data which can be exploited for 
compression purposes.  We carried out assessments with the Lempel-Ziv 
run length encoding compression algorithm, which is used in the gzip command
in Unix.  

The data array of Figure \ref{figsim2} was of size 
1,923,840 bytes.  A random row/column permuted version of this array 
was of identical size.  This random
row/column permuting used uniformly distributed
ranks.  In all cases considered here, 32-bit floating point storage 
was used, and each array, stored as an  image, 
contained a small additional ascii header.

Applying Lempel-Ziv compression to these data tables yielded gzip-compressed
files of size, respectively, 1,720,368 and 1,721,284 bytes.  There is not 
a great deal of compressibility present here.  The row/column permutation,
as expected, is not of help. 

Next, we used our hierarchical wavelet transform, where the output is of 
exactly the same dimensions as our input (here: $1200 \times 400$).  Again 
we stored this transformed data in an image format, using 32-bit floating 
point storage.  So the size of the transformed data was again exactly 
1,923,840 bytes.  This is irrespective of any wavelet filtering (by setting
wavelet transform coefficients to 0), and is solely due to the file size, 
based on so much data, each value of which is stored as a 32-bit floating 
point value.  

Then we compressed, using Lempel-Ziv, 
the wavelet transform data.  We looked at three 
alternatives: no wavelet filtering used; wavelet filtering such that 
coefficients with value up to 0.05 were set to 0; and wavelet filtering such 
that coefficients with value up to 0.1 were set to 0.   

The Lempel-Ziv compressed files were respectively of sizes: 
1,780,912, 1,435,477, and 1,108,743 bytes.  We see very clearly that Lempel-Ziv
run length encoding is benefiting from our wavelet filtering, provided
that there is sufficient filtering. 

It is useful to proceed further with this study, to see if simple 
filtering, similar to what we are doing in wavelet space, can be applied
also to the original data (or, for that matter, the permuted data).

\begin{table}
\begin{tabular}{|llllll|}\hline
HWT thresh. &   MSE &   Comp. size &  Orig. thresh. & MSE &  Comp. size \\ \hline
0      &  0 &  1,780,912   &   0    &   0   &   1,720,368  \\
0.05   &  0.0032 &  1,435,477  &  0.25  &  0.0028  &  1,447,984  \\
0.1    &  0.0119 &  1,108,743  &  0.39  &  0.0111  &  1,252,793  \\ \hline
\end{tabular}
\caption{Columns from left to right: Haar wavelet transform threshold 
applied to transformed data, with values lower than or equal to this set to 0; 
mean square error between input data and reconstructed, filtered data; 
compression size in bytes of the reconstructed, filtered data; threshold 
applied to the original data, with values lower than or equal to this 
set to 0; mean square error between input data and thresholded data; 
compression size in bytes of the thresholded data.  See text for 
discussion.}
\label{tabxx}
\end{table}

Table \ref{tabxx} shows what we found.  The mean square error (MSE) here
is $\sum_{ij} (\hat{x}_{ij} - x_{ij})^2 / \sum_{ij} x_{ij}^2$.  What is 
noticeable is that in the 2nd and 3rd rows of the table processing the 
original data is seen to have 
{\em better} MSE coupled with {\em worse} compressibility.
{\em A fortiori} we claim that for the {\em same} MSE we would have 
even {\em worse} compressibility.  

Our wavelet approach therefore has performed better in this study than 
direct processing of the given data.  Furthermore, relative to arbitrary
permuting of the rows and columns, our given data represents a relatively 
favorable case.  In practice, we have no guarantee that rows and columns
are arranged such that like values are contiguous.  But by design we have
such a favorable situation here.  

In concluding this study, we note the following work which uses a Fourier
transform, rather than a wavelet transform, on a hierarchy or rooted tree.
Kargupta and Park (2004) study the situation when a decision tree is 
given to begin with, and apply a Fourier transform to expedite 
compression.  For our purposes, a hierarchical tree is determined in 
order  to structure 
the data, and comprises 
the first part of a data processing pipeline.  A subsequent
part of the processing pipeline is the wavelet transform of the
hierarchically structured data.

\section{Wavelet-Based Multiway Tree Building:
Approximating a Hierarchy by Tree Condensation}
\label{sectcollapse}

Deriving a partition from a hierarchical clustering has always been a crucial 
task in data analysis (an early example is Mojena, 1977).   In this section
we will explore a new approach to deriving a partition from a dendrogram.  
Deriving a partition from a dendrogram is equivalent to ``collapsing'' a 
strictly binary tree into a non-binary or multiway tree.  Now, concept 
hierarchies are increasingly used to expedite search in information 
retrieval, especially in cross-disciplinary domains where a specification 
of the various terminologies used can be helpful to the user.  For concept
hierarchies, a non-binary or multiway tree is preferred for this purpose,
and may, in practice, be determined from a binary tree (as do Chuang and Chien,
2005).  

It is interesting to note that Lerman (1981, pp.\ 298--299) 
addresses this same issue of the ``condensation of a tree to the levels
corresponding to its significant nodes'', and then proceeds to discuss
the global criterion used (for instance, in an agglomerative hierarchical 
algorithm) vis-\`a-vis a local criterion (where the latter is used to 
judge whether a node is significant or not).  There is a clear parallel 
between this way of viewing the ``collapsing clusters'' problem and our 
way of tackling it.

A binary rooted tree, $H$, on $n$ observations has precisely $n-1$ levels;
 or $H$ contains precisely $n-1$ subsets of the set of $n$ observations.  The
interpretation of a hierarchical clustering often is carried out by cutting
the tree to yield any one of the $n-1$ possible partitions.  Our hierarchical
Haar wavelet transform affords us a neat way to approximate $H$ using a
smaller number of possible partitions.

Consider the detail vector at any given level: e.g., as
exemplified in Table \ref{table6}.  Any such detail
vector is associated with (i) a node of the binary tree; (ii) the level
or height index of that node; and (iii) a cluster,
or subset of the observation set.
With the goal of ``collapsing'' clusters, i.e.\ removing clusters that
are not unduly valuable for interpretation, we will impose a hard threshold
on each detail vector:  
{\em If the norm of the detail vector is less than a user-specified threshold,
then set all values of the detail vector to zero.}

Other rules could be chosen, in particular rules related directly to the
agglomerative clustering criterion used, but our choice of the thresholded
norm is a reasonable one.  
Our norm-based rule is not directly
related to the agglomerative criterion for the following reasons: (i) we
seek a generic interpretative aid, rather than an optimal but
criterion-specific rule; (ii) an optimal, criterion-specific rule would in
any case be best addressed by studying the overall optimality measure rather
than availing of the stepwise suboptimal hierarchical clustering; and (iii)
from naturally occurring hierarchies, as occur in very high dimensional
spaces (cf.\ Murtagh, 2004),
the issue of an agglomerative criterion is not important.

Following use of the norm-based cluster collapsing rule, the representation
of the reconstructed hierarchy is straightforward: the hierarchy's level
index is adjusted so that the {\em previous} level index additionally takes
the place of the given level index.  Examples discussed below
will exemplify this.

Properties of the approach include the following:

\begin{enumerate}
\item Rather than misleading increase in agglomerative clustering value or
level, we examine instead clusters (or nodes in the hierarchy).
\item This implies that we explore a cluster at a time, rather than a
partition at a time.  So the resulting retained clusters may well come from
different original partitions.
\item We take a strictly binary (2-way, agglomeratively constructed) tree
as input and determine a simplified, multiway tree as output.
\item A single scalar value filtering threshold --  a user-set parameter --
is used to derive this output, simplified, multiway tree from the input binary
tree.
\item The filtering is carried out on the wavelet-transformed tree; and
then the output, simplified tree is reconstructed from the wavelet transform
values.
\item The filtering is carried out on each node (in wavelet space) in
sequence.  Hence the computational complexity is linear.
\item Upstream of the wavelet transform, and hierarchical clustering,
we use correspondence analysis
to take frequency of occurrence data input, apply appropriate normalization,
and map the data of interest into an (unweighted) Euclidean space.
(See Murtagh, 2005.)
\item Again upstream of the wavelet transform, for
the binary tree we use minimal variance hierarchical clustering.
This agglomerative criterion favors compact clusters.
\item Our hierarchical clustering accommodates weights on the input
observables to be clustered.  Based on the normalization used in the
correspondence analysis, by design these weights here are constant.
\end{enumerate}

\subsection{Properties of Derived Partition}

\begin{figure*}
\begin{center}
\includegraphics[width=14cm]{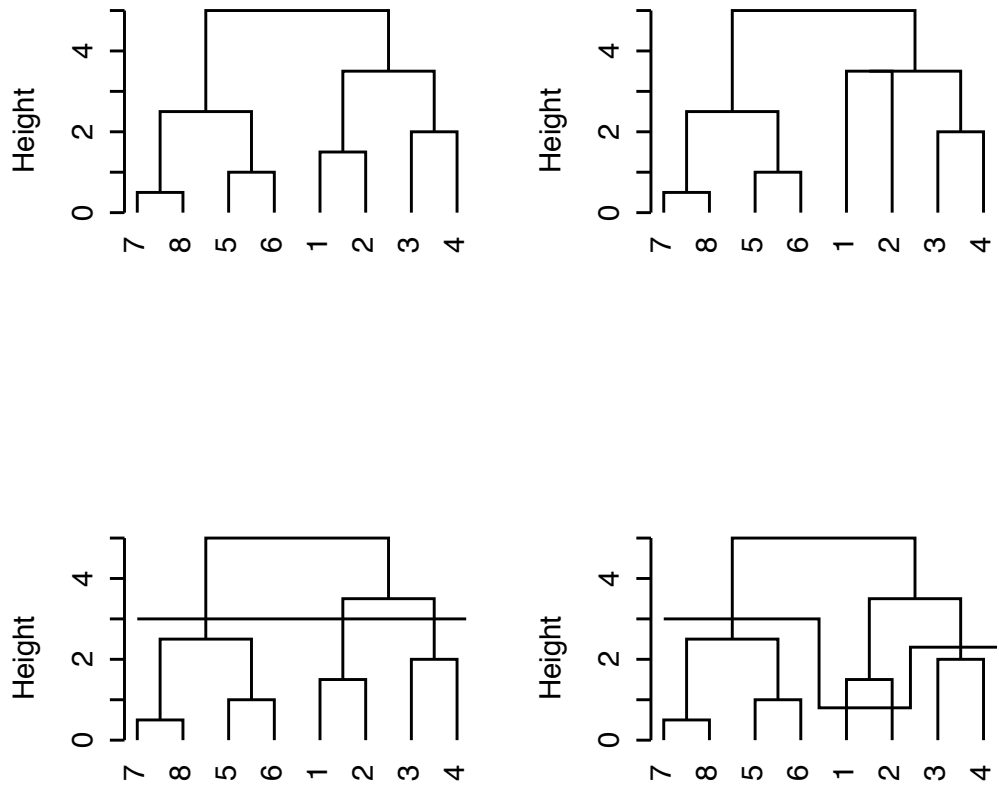}
\end{center}
\caption{Upper left: original dendrogram.  Upper right, multiway tree
arising from one collapsed cluster or node.  Lower left: a partition
derived from the dendrogram (see text for discussion).  Lower right:
corresponding partition for the multiway tree.}
\label{fig777}
\end{figure*}

A partition by definition is a set of clusters (sets) such that none are
overlapping, and their union is the global set considered.  So in Figure
\ref{fig777} the upper left hierarchy is cut, and shown  in the lower left,
 to yield
the partition consisting of clusters $(7, 8, 5, 6), (1, 2)$ and $(3, 4)$.
Traditionally, deriving such a partition for further exploitation is a
common use of hierarchical clustering.  Clearly the partition corresponds
to a height or agglomeration threshold.

In a multiway hierarchy, such as the one shown in the top right panel in
Figure \ref{fig777}, consider the same straight line drawn from left to
right, at approximately the same height or agglomeration threshold.  It
is easily seen that such a partition is the same as that represented by
the non-straight curve of the lower right panel.

From this illustrative example, we draw two conclusions: (i) in the case
of a multiway tree a partition is furnished by a horizontal cut of the
multiway tree -- accomplished exactly as in the case of the strictly
binary tree; and (ii) this horizontal cut of a multiway tree is identical
to a nonlinear curve of the strictly binary tree.  We can validly term
the nonlinear curve a piecewise horizontal one.

Note that the nonlinear curve used in Figure \ref{fig777}, lower right
panel, has nothing whatsoever to do with nonlinear cluster separation (in
any ambient space in which the clusters are embedded), nor with nonlinear
mapping.

\subsection{Implementation and Evaluation}
\label{sectimev}

We took Aristotle's {\em Categories} (see Aristotle, 350BC; Murtagh, 
2005) in English
containing 14,483
individual words.  We broke up the text into 24 files, in order to study
the sequential properties of the argument developed in this short
philosophical work.  In these 24 files, there were 1269 unique words.  We
selected 66 nouns of particular interest.  With frequencies of occurrence
in parentheses we had (sample only): man (104), contrary (72), same (71),
subject (60), substance (58), species (54), knowledge (50), qualities (47),
etc.  No stemming or other
preprocessing was applied on the grounds that singular and plurals could
well indicate different semantic content; cf.\ generic ``quantity'' versus
the set of specific, particular ``quantities''.  

The terms $\times$ subtexts data array was {\em doubled} (Murtagh, 2005)
 to produce
a $66 \times 48$ array: for each subtext $j$ with term frequencies of
occurrence $a_{ij}$, frequencies from a ``virtual subtext'' were
defined as $a^\prime_{ij} =
\mbox{max}_{ij} \ a_{ij} - a_{ij}$.  In this way the mass of term $i$, defined
as proportional to the associated row sum, is constant.  Thus what we have
achieved is to weight all terms identically.  (We note in passing that
{\em term} vectors therefore cannot be of zero mass.)

A correspondence analysis was carried out on the $66 \times 48$
table of frequencies with the aim of taking the set of 66 nouns endowed
with the $\chi^2$ metric (i.e., a weighted Euclidean distance between
{\em profiles}; the weighting is defined by the inverse subtext
frequencies) into a factor space endowed with the (unweighted)
Euclidean metric.  (We note in passing that any {\em subtexts}
of zero mass must
be removed from the analysis beforehand; otherwise inverse subtext frequency
cannot be calculated.)  Correspondence analysis provides a convenient and
general way to ``euclideanize'' the data, and any alternative could be 
considered also (e.g., as discussed in section 5.1 of Heiser, 2004).
A hierarchical clustering (minimum variance method) was
carried out on the factor coordinates of the 66 nouns.  Such a hierarchical
clustering is a strictly binary (i.e.\ 2-way), rooted tree.

\begin{figure*}
\begin{center}
\includegraphics[width=14cm]{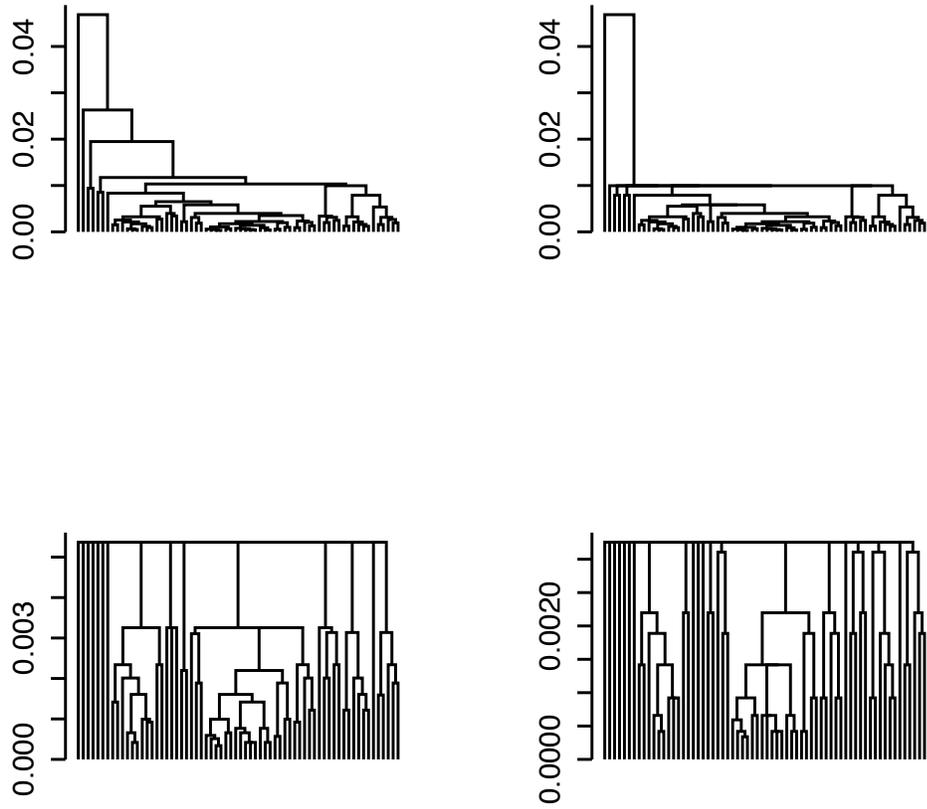}
\end{center}
\caption{Upper left: original hierarchy.  Upper right, lower left,
and lower right show increasing approximations to the original hierarchy
based on the ``cluster collapsing'' approach described.}
\label{fig7}
\end{figure*}

The norms of detail vectors had minimum, median and maximum values as
follows: 0.0758, 0.2440 and 0.6326, and these influenced the choice of
threshold.  Applying thresholds of 0, 0.2, 0.3 and 0.4 gave rise to the
following numbers of ``collapsed'' clusters with, in brackets, the
mean squared error between approximated data and original input data: 0
(0.0), 23 (0.0054),  44 (0.0147),  and 55 (0.0164).   Figure \ref{fig7}
shows the corresponding reconstructed and approximated hierarchies.

In the case of the threshold 0.3 (lower left in Figure \ref{fig7})
we have noted that 44 clusters were collapsed, leaving just 21 partitions.
As stated the objective here is precisely to approximate the dendrogram
output data structure in order to facilitate further study and interpretation
of these partitions.

\begin{figure*}
\begin{center}
\includegraphics[width=14cm]{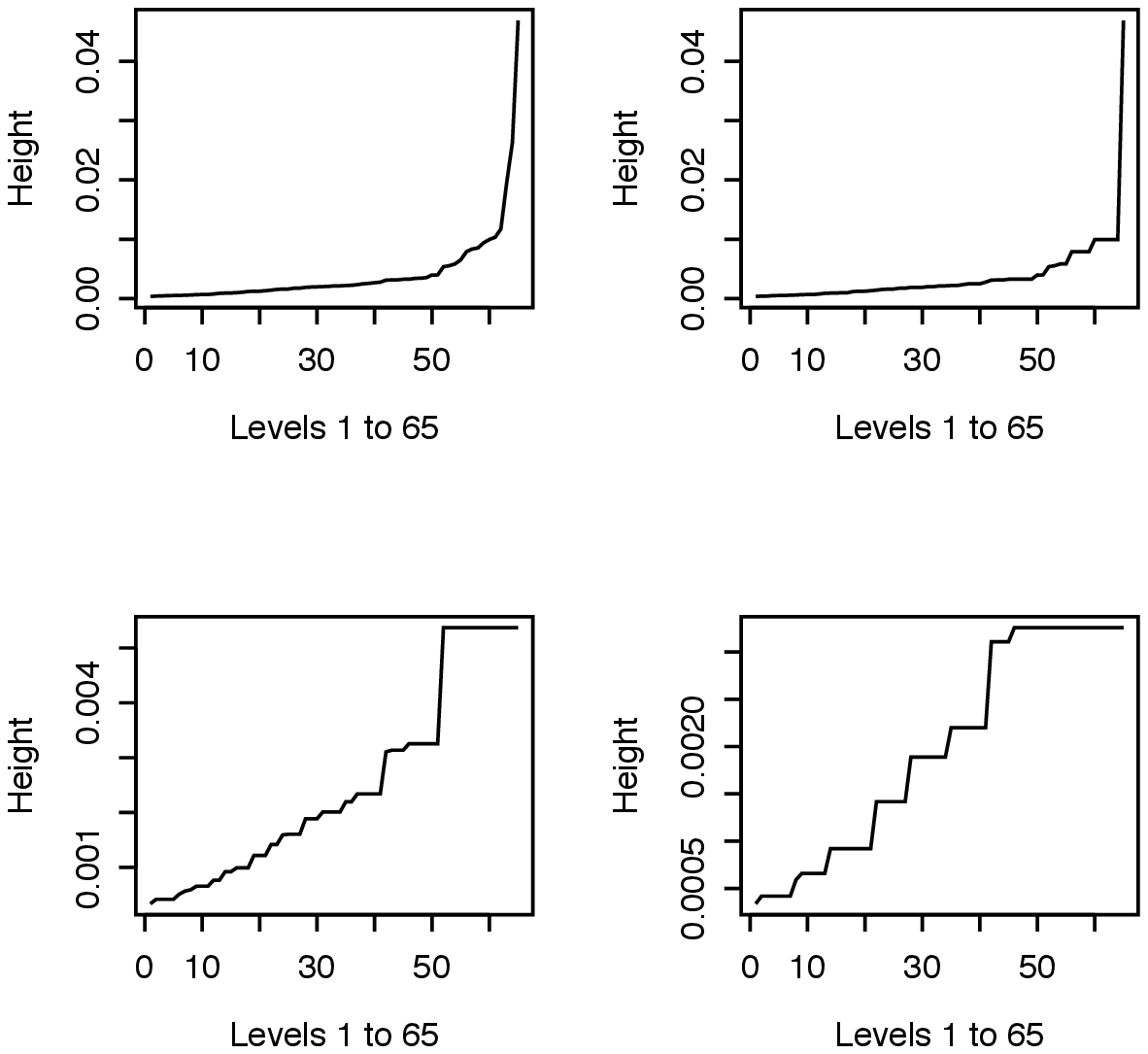}
\end{center}
\caption{Agglomerative clustering levels (or heights) for each of the
hierarchies shown in Figure \ref{fig7}.}
\label{fig8}
\end{figure*}

\begin{figure*}
\begin{center}
\includegraphics[trim=0mm 60mm 0mm 0mm,width=14cm,clip=true]{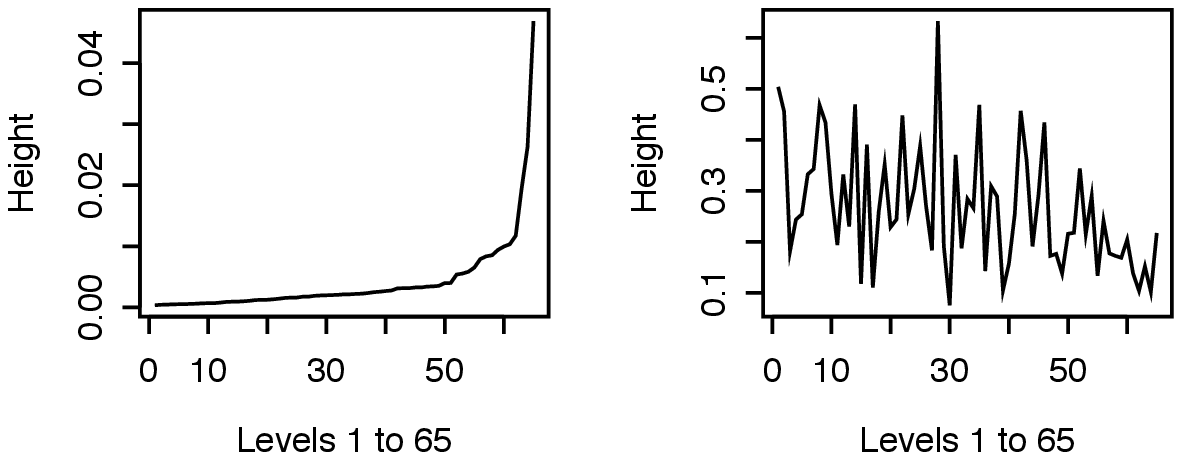}
\end{center}
\caption{Agglomerative levels (left; as upper left in Figure \ref{fig8},
and corresponding to the original -- upper left --
hierarchy shown in Figure \ref{fig7}),
and detail norms (right), for the hierarchy used.  Detail norms are
used by us as the basis for ``collapsing clusters''.}
\label{fig9}
\end{figure*}

Figure \ref{fig8} shows the sequence of agglomerative levels where each
panel corresponds to the respective panel in Figure \ref{fig7}.  It is
clear here why these agglomerative
levels are very problematic if used for choosing a
good partition: they increase with agglomeration, simply because the cluster
centers are getting more and more spread out as the sequence of agglomerations
proceeds.  Directly using these agglomerative levels has been a way
to derive a partition for a very long time (Mojena, 1977).  
To see how the detail norms used by us here are different, see
Figure \ref{fig9}.

\subsection{Collapsing Clusters Based on Detail Norms: Evaluation
Vis-\`a-vis Direct Partitioning}

Our cluster collapsing algorithm is: wavelet-transform the hierarchical
clustering; for clusters corresponding to detail norm less than a set
threshold, set the detail norm to zero, and the corresponding increase in
level in the hierarchy also; reconstruct the hierarchy.  We look at a range
of threshold values.  To begin with, the hierarchical clustering is strictly
binary.  Reconstructed hierarchies are multiway.

For each unique level of such a multiway hierarchy (cf.\ Figure \ref{fig7})
how good are the partitions relative to a direct, optimization-based
alternative?  We use the algorithm of Hartigan and Wong (1979)  with a
requested number of clusters in the partition given by the same number of
clusters in the collapsed cluster multiway hierarchy.  In regard to the
latter, we look at all unique partitions. (In regard to initialization
and convergence criteria, the Hartigan and Wong algorithm
implementation in the R package, www.r-project.org, was used.)

We characterize partitions using the average cluster variance, \\
$ 1/|Q| \sum_{i \in q; q \in Q} 1/|q| \| i - q \|^2$.
Alternatively we assessed the sum of squares: $\sum_{q \in Q}
\| i - q \|^2$.  Here, $Q$ is partition, $q$ is a cluster, and
$\| i - q \|^2$ is Euclidean distance squared between a vector $i$ and its
cluster center $q$.  (Note that $q$ refers both to a set and to a cluster
center -- a vector -- here.)  Although this is a sum of squares criterion,
as Sp\"ath (1985, p.\ 17) indicates, it is on occasion
(confusingly) termed the variance criterion.
In either case, we target compact clusters with this k-means clustering
algorithm,  which is also the target of
our hierarchical agglomerative clustering algorithm.
A k-means algorithm aims to optimize the criterion, in the Sp\"ath sense,
 directly.

In Table \ref{tab3} we see that the partitions of our multiway hierarchy
are about half as good as k-means in terms of overall compactness
(cf.\ columns 3 and 5).  Close
inspection of properties of clusters in different partitions indicated why
this was so: with a poor or low compactness for one cluster very early on
in the agglomerative sequence,
the stepwise algorithm used by the multiway hierarchy had to live with this
cluster through all later agglomerations; and the biggest sized cluster
(i.e.\ largest cluster cardinality) in the stepwise agglomerative tended to
be a little bigger than the biggest sized cluster in the k-means result.

This is an acceptable result: after all,
k-means optimizes this criterion directly.
Furthermore, the
multiway hierarchy preserves embededness relationships which are not
necessarily present in any sequence of results ensuing from a k-means
algorithm.  Finally, it is well-known that seeking to directly optimize
a criterion such as k-means will lead to a better outcome than the stepwise
refinement used in the stepwise agglomerative algorithm.

If we ask whether
k-means can be applied once, and then k-means applied to individual clusters
in a recursive way, the answer is of course affirmative -- subject to
prior knowledge of the number of levels and the value of k throughout.  It
is precisely in such areas that our hierarchical approach is to be preferred:
we require less prior knowledge of our data, and we are satisfied with
the downside of
global approximate fidelity between output structure and our data.

\begin{table}
\setlength{\tabcolsep}{0.5mm}
\begin{center}
\begin{tabular}{rrrrrr}\hline
Agglom.  &   Multiway tree   & Multiway tree & Partition   &  K-means     & \\
level    &   height          & partition SS  & cardinality & partition SS & \\
\hline
1        &   0.00034         &   0.095       &   65        & 0.062 & \\
2        &   0.00042         &   0.229       &   61        & 0.091 & \\
3        &   0.00051         &   0.340       &   60        & 0.146 & \\
4        &  0.00057          &   0.397       &   59        & 0.205 & \\
5        &  0.00059          &   0.485       &   58        & 0.156 & \\
6        &   0.00066         &   0.739       &   55        & 0.239 & \\
7        &   0.00077         &   1.115       &   53        & 0.347 & \\
8        &   0.00092         &   1.447       &   51        & 0.484 & \\
9        &   0.00099         &   1.723       &   48        & 0.582 & \\
10       &   0.00122         &   2.329       &   45        & 0.852 & \\
11       &   0.00142         &   2.684       &   43        & 0.762 & \\
12       &   0.00159         &   3.101       &   42        & 0.865 & \\
13       &   0.00161         &   3.498       &   39        & 1.161 & \\
14       &   0.00189         &   3.938       &   36        & 1.354 & \\
15       &   0.00201         &   4.954       &   32        & 1.873 & \\
16       &   0.00220         &   5.293       &   30        & 2.178 & \\
17       &   0.00234         &   6.957       &   25        & 3.007 & \\
18       &   0.00311         &   7.204       &   24        & 2.722 &  \\
19       &   0.00314         &   8.627       &   21        & 3.497 & \\
20       &   0.00326         &  10.192       &   15        & 5.426 & \\
21       &   0.00537         &  18.287       &    1        & 18.287  &  \\
\hline
\end{tabular}
\end{center}
\caption{Analysis of the unique partitions in the multiway tree shown
on the lower left of Figure \ref{fig7}.  Partitions are benchmarked using
k-means to construct partitions where k is the same value as found in the
multiway tree.  SS = sum of squares criterion value.}
\label{tab3}
\end{table}

\section{Wavelet Decomposition: Linkages with  Approximation and Computability}
\label{sect9}

A further application of wavelet-transformed dendrograms is currently 
under investigation and will be briefly described here.  

In domain theory (see Edalat, 1997; 2003), 
a Scott  model considers a computer program
as a function from one (input)  domain to another (output) domain.  
If this function is continuous then the computation is well-defined and 
feasible, and the output is said to be computable.  
The Scott model is concerned with real number computation, or 
computer graphics programming where, e.g., object overlap may be true,
false, or unknown and hence best modeled with a partially ordered set.  
In the Scott model, well-behaved approximation can benefit therefore from
function monotonicity.  

An alternative, although closely related, structure with which domains
are endowed is that of spherically complete ultrametric spaces.   The 
motivation comes from logic programming, where non-monotonicity may well 
be relevant (this arises, for example, with the negation operator).  Trees
can easily represent positive and negative assertions.  The general notion
of convergence, now, is related to {\em spherical completeness} (Schikhof,
1984; Hitzler and Seda, 2002).  If we have any set of embedded clusters,
or any chain, $q_k$, then the condition that such a chain be non-empty,
$\bigcap_k q_k \neq \emptyset$, means that this ultrametric space is 
non-empty.  This gives us both a concept of completeness, and also a 
fixed point which is associated with the ``best approximation'' of the 
chain.  

Consider our space of observations, $X = \{ x_i | i \in I \}$.  The 
hierarchy, $H$, or binary rooted tree, defines an ultrametric space.  For 
each observation $x_i$, by considering the chain from root cluster to 
the observation, we see that $H$ is a spherically complete ultrametric
space.

Our wavelet transform allows us to read off the chains that make the 
ultrametric space a spherically complete one.   A non-deterministic 
worst case $O(n)$ data re-creation algorithm ensues, compared to a more
usual non-deterministic worst case $O(m)$ data recreation algorithm.  The 
importance of this result is when $m >> n$.  

In our current work (Murtagh, 2007) we are studying this perspective 
based on (i) a body of texts from the same author, and (ii) a library of 
face images.

\section{Conclusion}

We have described the theory and practice of 
a novel wavelet transform, that is based on an available 
hierarchic clustering of the data table. We have generally 
used Ward's 
minimum variance agglomerative hierarchical clustering in this work. 

We have described this new method through a number of examples, both
to illustrate its properties and to show its operational use.

A number of innovative applications were undertaken with this new
approach.  These lead to various exciting open possibilities in regard
to data mining, in particular in high dimensional spaces.

\section*{Acknowledgements}

Dimitri Zervas converted the hierarchical clustering and new Haar wavelet
transform into C/C++ from the author's R and Java codes.  

\section*{References}

\medskip
\noindent
ALTAISKY, M.V. (2004).  ``p-Adic Wavelet Transform and Quantum Physics'',  
{\em Proc. Steklov Institute of Mathematics}, vol. 245, 34--39.

\medskip
\noindent
ALTAISKY, M.V. (2005).  {\em Wavelets: Theory, Applications, Implementation}, 
Universities Press.

\medskip
\noindent
ARISTOTLE (350 BC).  {\em The Categories}.  Translated by E.M. Edghill.
Project Gutenberg e-text, www.gutenberg.net

\medskip
\noindent
BENEDETTO,  R.L. (2004).  ``Examples of Wavelets for Local Fields'',
In C. Heil, P. Jorgensen, D. Larson, eds., {\em Wavelets, Frames, and
Operator Theory}, {\em Contemporary
Mathematics Vol.\ 345}, 27--47.

\medskip
\noindent
BENEDETTO, J.J. and BENEDETTO, R.L. (2004). ``A Wavelet Theory for Local
Fields and Related Groups'', {\em The Journal of Geometric Analysis},
14, 423--456.

\medskip
\noindent
BENZ\'ECRI,  J.P. (1979). 
{\em La Taxinomie}, 2nd ed., Paris: Dunod.

\medskip
\noindent
CHAKRABARTI, K.,  GAROFALAKIS, M., RASTOGI, R. and SHIM, K. (2001). 
``Approximate Query Processing using Wavelets'', {\em VLDB Journal, 
International Journal on Very Large Databases},
10, 199--223.  

\medskip
\noindent
SHUI-LUNG CHUANG and LEE-FENG CHIEN (2005).  ``Taxonomy Generation for 
Text Segments: A Practical Web-Based Approach'', {\em ACM Transactions 
on Information Systems}, 23, 363--396.

\medskip
\noindent
DEBNATH, L. and MIKUSI\'NSKI, P. (1999).
{\em Introduction to Hilbert Spaces with Applications}, 2nd edn.,
Academic Press.

\medskip
\noindent
 EDALAT, A. (1997).  ``Domains for Computation in Mathematics, 
Physics and Exact Real Arithmetic'', 
{\em Bulletin of Symbolic Logic}, 3, 401--452.

\medskip
\noindent
EDALAT, A. (2003).  ``Domain Theory and Continuous Data Types'', 
lecture notes, www.doc.ic.ac.uk/$\sim$ae/teaching.html

\medskip
\noindent
FISHER,  R.A. (1936).  ``The Use of Multiple Measurements in Taxonomic
Problems'', {\em The Annals of Eugenics}, 7, 179--188.

\medskip
\noindent
FOOTE, R.,  MIRCHANDANI, G.,  ROCKMORE, D.,  HEALY, D. and
OLSON, T. (2000a).  ``A Wreath Product Group Approach to Signal and
Image Processing: Part I -- Multiresolution Analysis'',
{\em IEEE Transactions on Signal Processing}, 48, 102--132

\medskip
\noindent
FOOTE, R.,  MIRCHANDANI, G.,  ROCKMORE, D., HEALY, D. and OLSON, T.
(2000b). 
``A Wreath Product Group Approach to Signal and Image Processing:
Part II -- Convolution, Correlations and Applications'',
{\em IEEE Transactions on Signal Processing}, 48, 749--767.

\medskip
\noindent
FOOTE, R. (2005). ``An Algebraic Approach to Multiresolution Analysis'',
{\em Transactions of the American Mathematical Society}, 357, 5031--5050.

\medskip
\noindent
FRAZIER, M.W. (1999).  {\em An Introduction to Wavelets through Linear 
Algebra},
New York: Springer.

\medskip
\noindent
H\"ARDLE, W. (2000).
 {\em Wavelets, Approximation, and Statistical Applications},
Berlin: Springer.

\medskip
\noindent
HARTIGAN, J.A. and WONG, M.A. (1979).  ``A K-Means Clustering Algorithm'',
{\em Applied Statistics}, 28, 100--108.

\medskip
\noindent
HEISER, W.J. (2004).  ``Geometric Representation of Association between 
Categories'', {\em Psychometrika}, 69, 513--545.

\medskip
\noindent
HITZLER, P. and SEDA, A.K. (2002).  ``The Fixed-Point Theorems of Priess-Crampe
and Ribenboim in Logic Programming'', {\em 
Fields Institute Communications}, 32,
219--235.

\medskip
\noindent
JOE, M.J.,  WANG, K.-Y. and KIM, S.-W. (2001). ``Wavelet Transformation-Based
Management of Integrated Summary Data for Distributed Query 
Processing'', {\em Data and Knowledge Engineering}, 39,
293--312.

\medskip
\noindent
JOHNSON, S.C. (1967).
``Hierarchical Clustering Schemes'', {\em Psychometrika}, 32, 241--254. 

\medskip
\noindent
KARGUPTA, H. and PARK, B.-H. (2004). ``A Fourier Spectrum-Based Approach to 
Represent Decision Trees for Mining Data Streams in Mobile Environments'',
{\em IEEE Transactions on Knowledge and Data Engineering}, 16,
216--229.

\medskip
\noindent
KHRENNIKOV, A.Yu. and KOZYREV, S.V. (2006), ``Ultrametric Random 
Field'', http://arxiv.org/abs/math.PR/0603584 

\medskip
\noindent
KOZYREV, S.V. (2002). ``Wavelet Analysis as a p-Adic Spectral Analysis'',
{\em Math. Izv.}, 66, 367--376.  http://arxiv.org/abs/math-ph/0012019

\medskip
\noindent
KOZYREV, S.V. (2004).  ``P-Adic Pseudo-differential Operators and
p-Adic Wavelets'',
{\em Theoretical and Mathematical Physics}, 138, 322--332.

\medskip
\noindent
LERMAN, I.C. (1981). {\em Classification et Analyse Ordinale des 
Donn\'ees}, Dunod. 

\medskip
\noindent
MURTAGH, F. (1985). {\em Multidimensional Clustering Algorithms},
W\"urzburg: Physica-Verlag.

\medskip
\noindent
MURTAGH, F. (1998).
``Wedding the Wavelet Transform and Multivariate Data Analysis'', 
{\em Journal of Classification}, 15, 161--183.

\medskip
\noindent
MURTAGH, F.,  STARCK, J.-L. and BERRY, M. (2000).  
``Overcoming the Curse of Dimensionality in Clustering by Means of the 
Wavelet Transform'', 
{\em The Computer Journal}, 43, 107--120.

\medskip
\noindent
MURTAGH, F. (2004). ``On Ultrametricity, Data Coding, and Computation'',
{\em Journal of Classification}, 21, 167--184.

\medskip
\noindent
MURTAGH, F. (2005).
{\em Correspondence Analysis and Data Coding with Java and R},
Chapman and Hall.  

\medskip
\noindent
MURTAGH, F. (2006).  ``Haar Wavelet Transform of a Dendrogram: Additional
Notes'', \\
http://www.cs.rhul.ac.uk/home/fionn/papers/HWTden\_notes.pdf

\medskip
\noindent
MURTAGH, F. (2007).  ``On ultrametric algorithmic information'', 
in preparation.

\medskip
\noindent
NABBEN, R. and VARGA, R.S. (1994). ``A Linear Algebra Proof that the 
Inverse of a 
Strictly Ultrametric Matrix is a Strictly Diagonal Dominant Stieltjes
Matrix'', {\em SIAM Journal on 
Matrix Analysis and Applications}, 15, 107--113.

\medskip
\noindent
OCKERBLOOM, J.M. (2003).  {\em  Grimms' Fairy Tales}, \\ 
http://www.cs.cmu.edu/$\sim$spok/grimmtmp

\medskip
\noindent
SCHIKHOF, W.M. (1984).  {\em Ultrametric Calculus}, Cambridge University 
Press.

\medskip
\noindent
SP\"ATH, H. (1985).  {\em Cluster Dissection and Analysis}, Ellis Horwood.

\medskip
\noindent
STARCK. J.-L. and MURTAGH, F. (2006). 
{\em Astronomical Image and Data Analysis},
Heidelberg: Springer.  Chapter 9: ``Multiple Resolution in Data Storage and 
Retrieval''.  (1st edn., 2002.) 

\medskip
\noindent
STRANG, G. and NGUYEN, T. (1996). {\em Wavelets and Filter Banks},
Wellesley-Cambridge Press.

\medskip
\noindent
TAO LI, QI LI, SHENGHUO ZHU, and MITSUNORI OGIHARA (2002).
``A Survey on Wavelet 
Applications in Data Mining'', {\em SIGKDD Explorations}, 4, 49--68.
             
\medskip
\noindent
VITTER, J.S.   and WANG, M. (1999).   
``Approximate Computation of Multidimensional
Aggregates of Sparse Data using Wavelets'', in 
{\em Proceedings of the 
ACM SIGMOD International Conference on Management of Data}, 193--204.

\end{document}